\documentclass[%
reprint,
 amsmath,amssymb,
 aps,
]{revtex4-2}

\usepackage{graphicx}
\usepackage{dcolumn}
\usepackage{bm}

\usepackage{graphicx}
\usepackage{epstopdf}
\usepackage{algorithmic}
\ifpdf
  \DeclareGraphicsExtensions{.eps,.pdf,.png,.jpg}
\else
  \DeclareGraphicsExtensions{.eps}
\fi
\usepackage[caption=false]{subfig}
\usepackage{amsmath,amssymb,amsfonts}
\usepackage{verbatim}
\usepackage{xcolor}
\usepackage{multirow}
\newcommand{\bs}{\boldsymbol}

\newcommand{\rmi}{\mathrm{i}}

\begin{document}

\preprint{APS/123-PRL}

\title{A universal model for drag on a spherical bubble}

\author{Qiang Sun}
 \email{qiang.sun@rmit.edu.au}
 \affiliation{%
 Australian Research Council Centre of Excellence for Nanoscale BioPhotonics, School of Science, RMIT University, Melbourne, VIC 3001, Australia
}

\author{Evert Klaseboer}
 \email{evert@ihpc.a-star.edu.sg}
 \affiliation{%
 Institute of High Performance Computing, 1 Fusionopolis Way, Singapore, 138632, Singapore
}

\date{\today}

\begin{abstract}
A theoretical expression for the drag on a spherical bubble is derived for the entire range from very viscous to inertial flow conditions. It is based on a solution for only that part of the velocity profile that determines the drag. It is assumed the surface of the bubble has a zero tangential stress condition. 
Excellent agreement with a previously proposed empirical model by Mei et al. is obtained. This shows that a theoretical framework with relatively simple physics can still predict the terminal velocity of a spherical bubble accurately.  To the best of our knowledge, this is one of the few models in fluid dynamics to predict drag on an object for a range of Reynolds numbers that spans many orders of magnitude.
\end{abstract}

\maketitle


\section{Introduction}



Gas bubbles in liquids appear in many natural settings, for example bursting bubbles in the ocean~\cite{Atamanchuk2020} are believed to play an important role in oceanic - atmospheric mass transfer~\cite{Berny2020}, or in bubbles trapped under impacting raindrops~\cite{KlaseboerPRL2014}. They appear in many industrial applications as well, for example in chemical engineering bubble columns~\cite{Almeras2018}, cleaning~\cite{Li2022}, and water purification. 
To understand bubble behavior is thus of great importance not only from a scientific physics viewpoint but also from a practical application viewpoint. One of the key problems is to determine the drag force acting on a rising bubble which in turn determines its terminal rise velocity based on the balance between the buoyancy force and the drag force. 

However, in fluid mechanics, it is in general very difficult to obtain an analytical solution for real life problems. The governing Navier-Stokes equations are categorized as one of the most difficult equations to solve in physics. Due to the non-linearity of the Navier-Stokes equations, very soon instabilities, vortex shedding, turbulence, flow separation and other complex phenomena will occur. 

It is always advantageous to have a theoretical solution as we can learn more about the physics involved than from pure numerical simulations. Thus the focus of this work is: to provide an analytical solution for the drag coefficient of a spherical bubble with radius $a$ moving at a velocity $U_0$ in a fluid with viscosity $\mu$ and density $\rho$. A key parameter in most fluid dynamics problems is the so-called Reynolds number, here defined as $\text{Re}=2aU_0 \rho/\mu$. It is extremely hard to find an analytical drag model that is valid for a large range of $\text{Re}$ due to the complexity of the Navier-Stokes equations. In this work, we take on the challenge to derive a universal analytical model for the drag on a rising spherical bubble for all $\text{Re}$ numbers.

Tiny amounts of surfactant can have a profound influence on the behavior of the tangential surface mobility of the bubble \cite{KlaseboerPRL2019}. Only for very pure water, the surface of the bubble exhibits `free slip'. For such a bubble, rising at constant velocity, unlike for a solid particle, the wake is almost absent, since the boundary layer does not detach due to the zero tangential shear condition on the bubble surface. 

The drag force is usually expressed in terms of a drag coefficient $C_d$~\cite{BatchelorBook1967}. 
%
%
%
The (viscous) Stokes flow drag limit for a bubble is well known (see Clift et al.~\cite{bookClift}) to be $C_d=16/\text{Re}$.  On the other hand, the (inertial) high Reynolds number limit is much harder to obtain theoretically. Moore~\cite{Moore1959} in a first attempt got $C_d=32/\text{Re}$ in 1959. He obtained this by taking the potential flow solution around a sphere and simply neglecting the tangential shear stress. However, it was pointed out to him by Batchelor (author of the classic book ``Introduction to fluid dynamics''~\cite{BatchelorBook1967}) that this result was wrong, since if an energy dissipation balance is done over the whole fluid domain, Batchelor got $C_d=48/\text{Re}$. Moore~\cite{Moore1963} then realised that the same result was obtained earlier by Levich in 1949~\cite{Levich1949}. After that, Moore came up with a new theory involving boundary layers and obtained $C_d=48/\text{Re}$, however his theory exhibits a diverging pressure at the back of the bubble (see also~\cite{Joseph2004}), which is not very satisfactory from a physics point of view. The factor of 1/3 missing from Moore's first attempt (the difference between 48/Re and 32/Re) is sometimes referred to as `viscous pressure correction'~\cite{KangLeal1988,Joseph2004}. In our theory, we not only derive an analytical model for drag on a spherical bubble but also discover the origin of this `viscous pressure' that corresponds to exactly 1/3 of the total drag for all Reynolds numbers. 

\section{Theory}


The problem is solved as follows: we start with a more general unsteady harmonic solution and transform this to the frequency domain with time dependency $\exp (-i\omega t)$, with time $t$ and angular frequency $\omega$, to obtain two equations: a Helmholtz equation and a Laplace equation with two constants, which are determined by the boundary conditions: the normal velocity and zero tangential stress condition. Here we use the fact that the drag on the bubble is only determined by the lowest order solution for the velocity (we thus ignore higher order terms). The steady case is the special situation of our theory at $t = 0$.

\begin{figure}[t]
\centering{}
\includegraphics[width=0.42\textwidth]{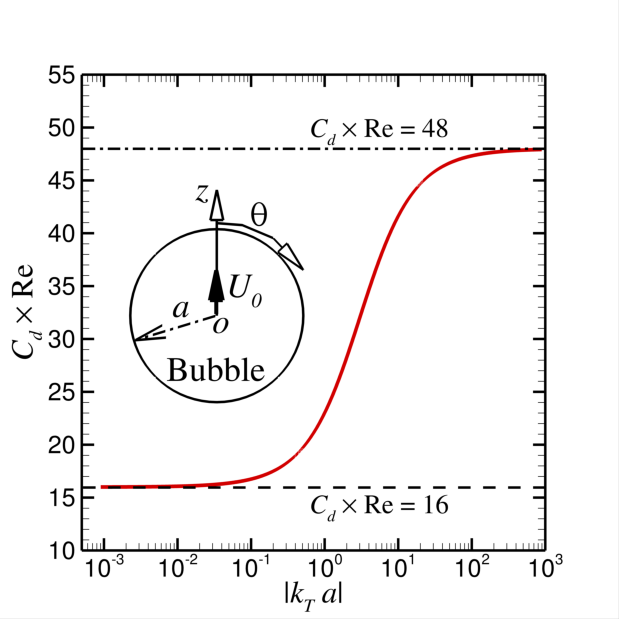}
\caption{Drag on a spherical bubble as a function of $|k_{T}a|$. The theoretical limits of viscous Stokes flow, $C_d \text{Re} = 16$, and inertial high Reynolds flow, $C_d \text{Re}= 48$, are recovered. A smoothly varying function is connecting these two limits.  }  \label{fig:cdRekTabubble}
\end{figure}

Recently, an analytical solution for an acoustic boundary layer around an oscillating rigid sphere was proposed by Klaseboer et al. \cite{KlaseboerPhysFl2020} based on the Nyborg~\cite{NyborgJASA1953} framework.  This framework describes acoustic and viscous waves in a compressible Newtonian fluid.  We use the same framework here for a rising bubble, with the difference that instead of no-slip boundary conditions, zero tangential stress is now imposed at the bubble surface and we use an incompressible framework. The governing equations for the  velocity, $\overline {\boldsymbol  v}(\boldsymbol x,t)$ and pressure, $\overline p(\boldsymbol x,t)$ are the Navier-Stokes equations
\begin{equation} \label{eq:NavierStokes}
\begin{aligned}
   \nabla \cdot  \overline {\boldsymbol  v} =  0 \quad ;\quad
  \rho\frac{\partial  \overline  {\boldsymbol  v} } {\partial t} + \rho \overline {\boldsymbol  v} \cdot \nabla \overline {\boldsymbol  v} =  - \nabla \overline p + \mu \nabla^2 \overline {\boldsymbol  v}.
\end{aligned}
\end{equation} 
Assume the velocity field $\overline {\boldsymbol v}(\boldsymbol x, t)$ is driven by a sphere of radius, $a$ executing small time harmonic motion with velocity: $\boldsymbol U_0 e^{-\rmi \omega t} = U_0 e^{-\rmi \omega t} \boldsymbol e_z$, along the $z$-direction and assuming the same harmonic time dependence for all quantities: $\overline p(\boldsymbol x, t) \sim p(\boldsymbol x) \exp (-\rmi \omega t)$ and $\overline {\boldsymbol v}(\boldsymbol x, t) \sim \boldsymbol u(\boldsymbol x) \exp (-\rmi \omega t)$, then Eq.~(\ref{eq:NavierStokes}) transforms into
\begin{equation} \label{eq:order_1_omega2}
\begin{aligned}
  \nabla \cdot \boldsymbol u  =  0 \quad ; \quad
   -\rmi \omega \rho \boldsymbol u    =  - \nabla p + \mu \nabla^2 \boldsymbol u.
\end{aligned}
\end{equation}
Or, alternatively written as:
\begin{equation}  \label{eq:NavierStokes2}
    \nabla^2 \boldsymbol u +k_T^2 \boldsymbol u -\frac{\nabla p}{\mu}= \boldsymbol 0 \quad ; \quad k_T^2  \equiv  \rmi \frac{\rho_0 \omega}{\mu}
\end{equation} 
with $k_T$  the (complex valued) transverse wave number. Now performing a Helmholtz decomposition as $\boldsymbol u = \boldsymbol u_T + \boldsymbol u_L$ with $\nabla \cdot \boldsymbol u_T = 0$ and $\nabla \times \boldsymbol u_L = \boldsymbol 0$ (thus we can define a potential function as $\boldsymbol u_L= \nabla \Phi$), we get
\begin{equation} 
\begin{aligned}\label{eq:Helmholtz}
&\nabla^2 \boldsymbol u_T + k_T^2 \boldsymbol u_T  = \boldsymbol 0 \quad ; \quad \nabla^2 \boldsymbol u_L = \boldsymbol 0 \quad \; \\
&\nabla p = \mu k_T^2 \boldsymbol u_L=\mu k_T^2 \nabla \Phi.
\end{aligned}
\end{equation}
Thus $p=\mu k_T^2 \Phi$. For an axial symmetric system:
\begin{equation}
    \begin{aligned} \label{eq:u_general2}
        \boldsymbol u =& \boldsymbol u_L + \boldsymbol u_T = \nabla \Phi - \nabla \times [\nabla \times (\boldsymbol x H)] = u_r \boldsymbol e_r + u_\theta \boldsymbol e_\theta\\ =& \quad U_0\left\{\quad -2\frac{h(r)}{r} \; \;+ \frac{d\phi(r)}{dr} \right\} \cos \theta \boldsymbol e_r \\ & +U_0\left\{\frac{1}{r}\frac{d}{dr}\Big[r h(r) \Big] -\frac{\phi(r)}{r}\right\} \sin \theta \boldsymbol e_\theta
    \end{aligned}
\end{equation}
with $\Phi(r,\theta) = U_0\phi(r) \cos \theta$ with $\nabla^2 \Phi  = 0$ and $H(r,\theta)=U_0 h(r) \cos \theta$ with $\nabla^2 H + k_T^2 H=0$.  $\theta$ is the angle the position vector makes with the vertical axis, $\boldsymbol e_\theta$ is the unit vector in the $\theta$ direction and $\boldsymbol e_r$ the unit vector in the $r$ direction. In Eq.~(\ref{eq:u_general2}) we have only used the lowest order Legendre polynomial solutions $P_1 (\cos \theta) = \cos \theta$, since these are sufficient to get the drag on the bubble (see Appendix). The functions $\phi$ and $h$ are defined up to a constant as $\phi(r)=-C_2  a^3/r^2$ and $h(r)=-C_1 \frac{a}{k_T}\frac{d}{dr} \left(\frac{\exp(ik_Tr)}{k_Tr}\right)$. The dimensionless constants $C_1$ and $C_2$ must be determined from the boundary conditions. The first boundary condition is that the normal velocity must satisfy $\boldsymbol u \cdot \boldsymbol e_r = U_0 \cos \theta$ to retain the spherical shape (the tangential velocity could be anything), this leads to
\begin{equation}
    \begin{aligned} \label{eq:VelBC2}
    -2 \frac{h}{r}+ \frac{d\phi}{dr} = 1 \quad ; \quad \text{Velocity condition at } r=a.
    \end{aligned}
\end{equation}
The vanishing tangential stress $\sigma_{r\theta}= \sigma_{\theta r} = 0 $ on the bubble surface leads to a second boundary condition at $r=a$:
\begin{equation}
    \begin{aligned} \label{eq:sigma_r2}
        \frac{\sigma_{r\theta}}{\mu}&=\frac{\partial u_\theta}{\partial r} - \frac{u_\theta}{r}+\frac{1}{r}\frac{\partial u_r}{\partial \theta} \\
        &= U_0 \sin \theta \left[ -\frac{2}{r} \frac{d\phi}{dr}+ 2 \frac{\phi}{r^2} + \frac{d^2h}{dr^2}\right]=0.
    \end{aligned}
\end{equation}
The two boundary conditions Eqs.~(\ref{eq:VelBC2}) and~(\ref{eq:sigma_r2}) will give a $2 \times 2$ matrix system that will determine the constants $C_1$ and $C_2$ as $C_1=3 \exp(-ik_Ta)/(ik_Ta -3)$ and $C_2=1/2 - 3(ik_Ta-1)/[k_T^2 a^2 (ik_Ta -3)]$. The radial stress can be written as:
\begin{equation}
    \begin{aligned} \label{eq:RadialStress2}
      \sigma_{rr}&= -p +2 \mu\frac{du_r}{dr}\\& = U_0 \mu\cos \theta \left[ -k_T^2 \phi +2\frac{d^2\phi}{dr^2} -4 \frac{d}{dr}\left( \frac{h}{r} \right)\right] \\
      &=\frac{U_0 \mu}{a} \cos \theta\left[\frac{k_T^2 a^2}{2} -9\left( 1 + \frac{2}{ik_Ta-3}\right) \right]\\
&= -9 \cos \theta\frac{U_0 \mu} {a} \left\{1 -\frac{6+\sqrt{2}|k_T a|}{9+3\sqrt{2}|k_T a|+ |k_T a|^2}\right \} \\
&\quad +\text{imaginary part}
\end{aligned}
\end{equation} 
where, in the last step, we have used the fact that $k_T^2a^2$ is a purely imaginary number. Also we used that $k_T$ has an equal real and imaginary part thus: $k_Ta=|k_Ta|/\sqrt{2} + i |k_Ta|/\sqrt{2}$. 
Exactly 2/3 of the contribution originates from the $2\frac{d^2\phi}{dr^2} -4 \frac{d}{dr}\left(\frac{h}{r}\right)$ term of Eq.~(\ref{eq:RadialStress2}) and 1/3 from the pressure term $ -k_T^2 \phi$ which thus corresponds to the `viscous pressure correction' \cite{KangLeal1988,Joseph2004}. Interestingly this ratio is valid for all values of $|kTa|$. The drag force is dependent on $\sigma_{rr}$ alone (since $\sigma_{r\theta}=0$) and becomes
\begin{equation}
    \begin{aligned} \label{eq:Fd2}
       F_d=\text{Real}\Big\{\int_0^\pi \sigma_{rr} \cos \theta \;2 \pi a \sin \theta \; a d\theta \Big\}.
    \end{aligned}
\end{equation}
 Entering Eq.~(\ref{eq:RadialStress2}) into Eq.~(\ref{eq:Fd2}), where we can use the identity $\int_0^\pi \cos^2 \theta \sin \theta d\theta = 2/3$, gives the drag coefficient $C_d$:
\begin{equation}
    \begin{aligned} \label{eq:Cd_kTa2}
        C_d = \frac{|F_d|}{\frac{1}{2}\rho U_0^2 \pi a^2} =\frac{48}{\text{Re}} \left\{1 -\frac{6+\sqrt{2}|k_T a|}{9+3\sqrt{2}|k_T a|+ |k_T a|^2}\right \}.
    \end{aligned}
\end{equation}
%
%
%
The classical limit for a thin boundary layer (corresponding to $|k_Ta|\gg 1$ or high $\text{Re}$ number), $C_d=48/\text{Re}$ is recovered. Also the Stokes flow limit (i.e $|k_Ta|\ll1$) for a free slip sphere is predicted as $C_d = 16/\text{Re}$ (Clift et al.~\cite{bookClift}); see also Fig.~\ref{fig:cdRekTabubble}. 

\begin{figure}[t]
\centering{}
\includegraphics[width=0.42\textwidth]{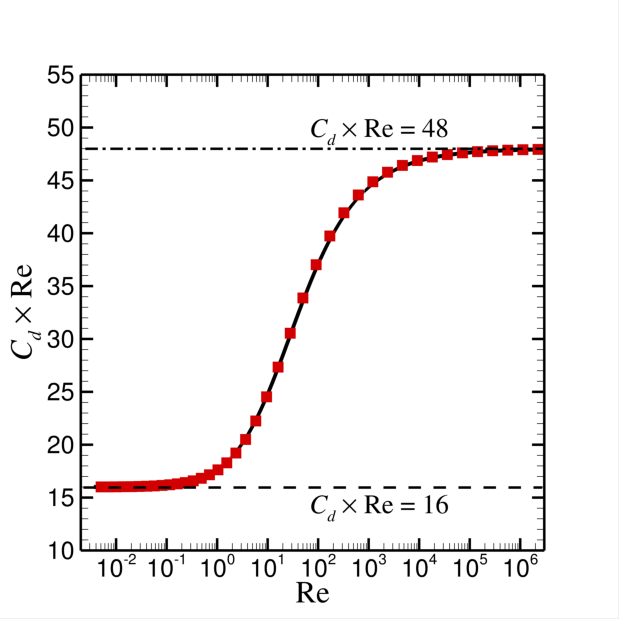}
\caption{Drag on a spherical bubble as a function of $\text{Re}$, line: our model of Eq.~(\ref{eq:Cd_Re}); dots: the empirical fit from Eq.~(\ref{eq:Cd_Magnaudet}). 
}  \label{fig:cdrebubble}
\end{figure}

Eq.~(\ref{eq:Cd_kTa2}) gives the drag coefficient as a function of $|k_Ta|$ without any fitting parameter, but it would be even more convenient to have it entirely as a function of Reynolds number. We can now distinguish two cases; For $\text{Re}\ll1$ we can construct a typical time scale from $\mu$, $U_0$ and $\rho$, such that it seems logical to take $\omega \sim \rho U_0^2/\mu$; a proportionality factor of exactly 2 seems to fit best, thus $|k_Ta| \sim \sqrt{\rho \omega/\mu}\; a = \rho U_0 a/\mu = \text{Re}/2$. Thus $\text{Re}=4|k_Ta|$. For $\text{Re}\gg1$ on the other hand we must go back to  Eq.~(\ref{eq:order_1_omega2}) and compare the terms $\rho \partial \overline  {\boldsymbol  v} / \partial t$ and $ \rho \overline {\boldsymbol  v} \cdot \nabla \overline {\boldsymbol  v}$. This will lead to $i \omega \rho U_0 \sim \rho U_0^2/L$. Now using $k_T^2=i \rho \omega/ \mu$ to get rid off $\omega$ we get $|k_T a|^2 \sim \rho U_0 a/\mu a/L = \text{Re} \;a/(2L)$. A typical length of $L=\sqrt{2}a$ seems to fit best. Thus $\text{Re}\sim 2 \sqrt{2}|k_Ta|^2 $ for $\text{Re}\gg1$. We finally propose to write the Reynolds number for intermediate $|k_Ta|$ cases as a combination of both expressions: $\text{Re}=4|k_Ta| + 2\sqrt{2}|k_Ta|^2$ or alternatively written:
\begin{equation}
    \sqrt{2} |k_Ta|=-1 + \sqrt{1 + \text{Re}/\sqrt{2}}.
\end{equation}
Using this in Eq.~(\ref{eq:Cd_kTa2}) gives the main result of this article:
\begin{equation}
    \begin{aligned} \label{eq:Cd_Re}
        C_d = \frac{48}{\text{Re}} \left\{1 -\frac{5+\sqrt{1 +\text{Re}/\sqrt{2}}}{7+2\sqrt{1+\text{Re}/\sqrt{2}}+ \text{Re}/(2\sqrt{2})}\right \}.
    \end{aligned}
\end{equation}
Eq.~(\ref{eq:Cd_Re}) represents the drag law for a bubble with a free slip surface, valid for all Reynolds numbers, obtained by a theoretical analysis. It gives excellent agreement with the empirical fit from Mei et al.~\cite{Mei1994} (see also Magnaudet and Eames~\cite{Magnaudet2000}):
\begin{equation}
    \begin{aligned} \label{eq:Cd_Magnaudet}
        C_d =\frac{16}{\text{Re}} \left\{1 +\left[\frac{8}{\text{Re}} +\frac{1}{2}\left(1 + \frac{3.315}{\sqrt{\text{Re}}} \right)\right]^{-1}\right \}
    \end{aligned}
\end{equation}
as is shown in Fig.~\ref{fig:cdrebubble}. Even though the functions in Eq.~(\ref{eq:Cd_Re}) and Eq.~(\ref{eq:Cd_Magnaudet}) do at first sight not resemble, yet the difference between the two equations is less than 1\% (with the maximum at $\text{Re} \approx$ 100).

\begin{figure}[t]
\centering{}
\includegraphics[width=0.42\textwidth]{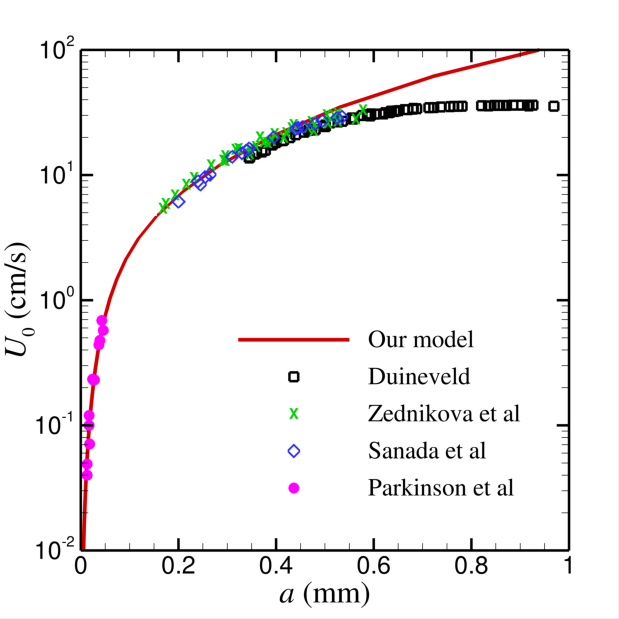}
\caption{Comparison with clean bubbles experiments; Parkinson et al~\cite{Parkinson2008} (very small bubbles), Zenikova et al.~\cite{Zednikova2010}, Sanada et al~\cite{Sanada2008} and Duineveld~\cite{Duineveld1995} (larger deformed bubbles).}  \label{fig:UaExp}
\end{figure}
%


In Fig.~\ref{fig:UaExp}, the terminal rise velocity, $U_0$, is plotted for a spherical bubble in water as a function of radius (by setting buoyancy equal to the drag force and calculate the velocity from there) for both our theory and some experimental data with clean bubbles \cite{Duineveld1995,Zednikova2010,Parkinson2008,Sanada2008}. Larger bubbles start to deform and become elliptic in shape such that the current theory becomes invalid.

 \section{Discussion and conclusion}

In this paper, we derived a universal analytical model for the drag on a rising bubble. In our model, the $48/\text{Re}$ high and the $16/\text{Re}$ low Reynolds limits were found without any fitting parameter. Also, we discovered the origin of the 1/3 viscous pressure correction, and this ratio is valid for all values of $\text{Re}$. Furthermore, unlike Moore's solution \cite{Moore1963,Joseph2004}, our solution does not blow up at the back of the bubble. For $\text{Re}>1$, we do not predict the (total) velocity field, only the part which contributes to the drag. 
 
From an application point of view, the high Re case is not that interesting for air bubbles in water (since deformation will occur and the bubble will become elliptic) or for very small bubbles in water, since surfactants will usually prevent the surface of the bubble to be entirely tangential stress free. The current theory is applicable for typical bubbles sizes of $a<0.5$ mm, of importance for numerous physical phenomena and industrial processes.
 
  
The classical potential flow pressure on the surface of the bubble, $p=(1- \sin^2 \theta \; 9/4) \rho U_0^2/2$, originating from the term $\rho\overline{\boldsymbol v}\cdot \nabla \overline{\boldsymbol v}$ in Eq.~(\ref{eq:NavierStokes}), is in fact of second order and does not contribute to the drag (as predicted by the d'Alembert paradox). It does, however, create an over-pressure at both poles and an under-pressure at the equator of the bubble, deforming the bubble into an ellipse if the surface tension is not large enough to keep the bubble spherical~\cite{KlaseboerEABE2011, Manica2016, Loth2008}. For air bubbles in water, the current theory is no longer valid for this reason for larger bubble radii (see also the larger Duineveld bubbles in Fig.~\ref{fig:UaExp}). 

To conclude, a theoretical solution was found for the drag on a spherical clean bubble for all Re numbers, with one fitting parameter of order one based on physical insight.  

\begin{acknowledgments}
Q.S. acknowledges the support from the Australian Research Council grants DE150100169, FT160100357 and CE140100003.
\end{acknowledgments}

\appendix
 
 \section {Neglect of higher order terms} \label{app:HigherOrderTerms}
 

Terms with higher order Legendre functions in Eq.~(\ref{eq:u_general2}) will not contribute to the drag, which can easily be seen by using  $\Phi_n(r,\theta) = \phi_n(r) P_n(\cos \theta)$ with $\nabla^2 \Phi_n = 0$ where $\phi_n(r) = C_{1,n}/r^{n+1}$ and $H_n(r,\theta)=  h_n(r) P_n(\cos \theta)$ with $\nabla^2 H_n + k_T^2 H_n=0$ where $h_n(r)= C_{2,n} h_n^{(1)}(k_Tr)$. Here, $n>1$, $h_n^{(1)}(k_Tr)$ is a spherical Hankel function of the first kind, and $C_{1,n},\,C_{2,n}$ are constants. Then,
\begin{widetext}
\begin{align} 
    \frac{u_r}{U_0a}&= \boldsymbol e_r \cdot \left\{ \nabla \left[ \phi_n(r) P_{n}(\cos{\theta})\right] + \nabla \times \nabla \times \left[-r h_n(r) P_{n}(\cos{\theta}) \bs{e}_{r}\right]\right\} \nonumber\\
    &= \boldsymbol e_r \cdot \left\{ \nabla \left[ \phi_n(r) P_{n}(\cos{\theta})\right] +  \nabla \times \left[ h_n(r) \frac{d}{d\theta}P_{n}(\cos{\theta}) \bs{e}_{\varphi}\right]\right \} \nonumber\\
    &= \boldsymbol e_r \cdot \bs{e}_r  \left[ \frac{d}{dr} \phi_n(r) P_{n}(\cos{\theta}) + \frac{1}{r} (n+1)n h_n(r) P_{n}(\cos{\theta}) \right] +\boldsymbol e_r \cdot \boldsymbol e_\theta .....\sim P_n(\cos \theta). \label{eq:HigherOrder}
\end{align} 
\end{widetext} 
In Eq.~(\ref{eq:HigherOrder}), we have used $\frac{1}{\sin{\theta}}\frac{d}{d\theta} \left[\sin{\theta}\frac{d}{d\theta}P_{n}(\cos{\theta})\right]  = -(n+1)n P_{n}(\cos{\theta})$. The term $du_r/dr\sim P_n(\cos \theta)$ in Eq.~(\ref{eq:RadialStress2}). 
Also, the pressure is proportional to $\Phi_n$ as  $p=\mu k_T^2 \Phi_n$.
Thus, all the terms in the radial stress (Eq.~\ref{eq:RadialStress2}) are proportional to $P_n(\cos \theta)$. Due to a property of Legendre functions: $\int_0^\pi P_n(\cos \theta) \cos \theta \sin \theta  d\theta=0$ for $n>1$, the drag force $F_d$ only depends on $P_1(\cos \theta) = \cos \theta$. The same conclusion was reached by Kang and Leal \cite{KangLeal1988}.

\bibliography{references}

\end{document}